\documentclass[conference]{IEEEtran}
\IEEEoverridecommandlockouts
\usepackage{cite}
\usepackage{amsmath,amssymb,amsfonts}
\usepackage{algorithmic}
\usepackage{graphicx}
\usepackage{textcomp}
\usepackage{xcolor}
\usepackage{multirow}
\usepackage{float}
\usepackage{url}
\usepackage{caption}
\floatstyle{plaintop}
\restylefloat{table}

\def\BibTeX{{\rm B\kern-.05em{\sc i\kern-.025em b}\kern-.08em
    T\kern-.1667em\lower.7ex\hbox{E}\kern-.125emX}}
\begin{document}

\title{Cyber Sentinel: Exploring Conversational Agents' Role in Streamlining Security Tasks with GPT-4
\thanks{This work was supported in part by the European Union’s Horizon 2020 research and innovation program under grant agreement No 101021911 (IDUNN Project).}
}

\author{\IEEEauthorblockN{1\textsuperscript{st} Mehrdad Kaheh }
\IEEEauthorblockA{\textit{Center for Ubiquitous Computing} \\
\textit{University of Oulu}\\
Oulu, Finland \\
Mehrdad.Kaheh@oulu.fi}
\and
\IEEEauthorblockN{2\textsuperscript{nd} Danial Khosh Kholgh}
\IEEEauthorblockA{\textit{Center for Ubiquitous Computing} \\
\textit{University of Oulu}\\
Oulu, Finland \\
Danial.Khoshkholgh@oulu.fi}
\and
\IEEEauthorblockN{3\textsuperscript{rd} Panos Kostakos}
\IEEEauthorblockA{\textit{Center for Ubiquitous Computing} \\
\textit{University of Oulu}\\
Oulu, Finland \\
Panos.Kostakos@oulu.fi}
}

\maketitle

\begin{abstract}
In an era where cyberspace is both a battleground and a backbone of modern society, the urgency of safeguarding digital assets against ever-evolving threats is paramount. This paper introduces \textit{Cyber Sentinel}, an innovative task-oriented cybersecurity dialogue system that is effectively capable of managing two core functions: explaining potential cyber threats within an organization to the user, and taking proactive/reactive security actions when instructed by the user. Cyber Sentinel embodies the fusion of artificial intelligence, cybersecurity domain expertise, and real-time data analysis to combat the multifaceted challenges posed by cyber adversaries. This article delves into the process of creating such a system and how it can interact with other components typically found in cybersecurity organizations. Our work is a novel approach to task-oriented dialogue systems, leveraging the power of chaining GPT-4 models combined with prompt engineering across all sub-tasks. We also highlight its pivotal role in enhancing cybersecurity communication and interaction, concluding that not only does this framework enhance the system's transparency (Explainable AI) but also streamlines the decision-making process and responding to threats (Actionable AI), therefore marking a significant advancement in the realm of cybersecurity communication.
\end{abstract}


\section{Introduction}
In the modern digital landscape, cybersecurity has emerged as a pivotal concern, as organizations and individuals alike are becoming increasingly reliant on interconnected technologies. The rapid proliferation of digital assets, cloud services, and the Internet of Things has brought about unprecedented conveniences but has also introduced intricate security challenges. The omnipresence of cyber threats, ranging from malware and phishing attacks to sophisticated hacking attempts, has underscored the need for robust defensive mechanisms that can adapt and respond to evolving tactics. In a digital world that is expanding each day in size, it is not reasonable to expect human experts to sift through millions of data files and logs, hoping to find possible threats on their own.

Over the years, artificial intelligence tools have provided significant contributions to various cybersecurity tasks, ranging from predictive use cases such as vulnerability analysis \cite{10.1145/3382025.3414952,8908835}, threat hunting \cite{9458828, 10224990b}, intrusion prediction \cite{Ansari2022,8993037} and intrusion detection \cite{sheikhi2022novel, 9527966,CHORAS2021705} to responsive measures like automated mitigation \cite{s19051114} and remediation \cite{9406576}. Generative AI (GenAI) \cite{cao2023comprehensive} has especially shined throughout the recent years with the emergence of Transformer-based Large Language Models (LLMs) such as BERT\cite {devlin2018bert} and GPT \cite{setianto2021gpt, radford2018improving}. The application of LLMs in cybersecurity remains relatively underexplored compared to other fields, especially considering their demonstrated efficacy in diverse domains.

In this paper, we aim to investigate the applicability of LLMs in a number of cybersecurity-related tasks. Specifically, we introduce a conversational agent called \emph{"Cyber Sentinel"}, backed by the OpenAI GPT-4 language model, and explore its usability in streamlining common security tasks. Cyber Sentinel's goal is to assist in querying and analyzing data from a Cyber Threat Intelligence (CTI) feed, such as those available in common Open Source Intelligence (OSINT) frameworks and feeds, and present this intelligence in a user-friendly format to a security operator for further processing. Additionally, Cyber Sentinel is designed to handle a number of security actions in response to a threat, such as updating firewall rules and updating SIEM configurations. The objectives are contextualized within the realms of explainable AI \cite{10.1007/978-3-030-32236-6_51} and actionable AI \cite{10.1145/3424690}, both of which have garnered significant attention in the recent AI literature \cite{linkov2020cybertrust}. In short, we aim to answer the following research questions:

\begin{itemize}
    \item \textbf{RQ1:} Can LLMs be utilized to understand cybersecurity logs, events, and threat feeds and explain them adequately to a human operator? \emph{(Explainable AI)} 
    \item \textbf{RQ2:} Can LLMs take security actions based on instructions from a human operator? \emph{(Actionable AI)}
\end{itemize}

\begin{figure*}[h!]
\centerline{\includegraphics[width=0.8\textwidth]{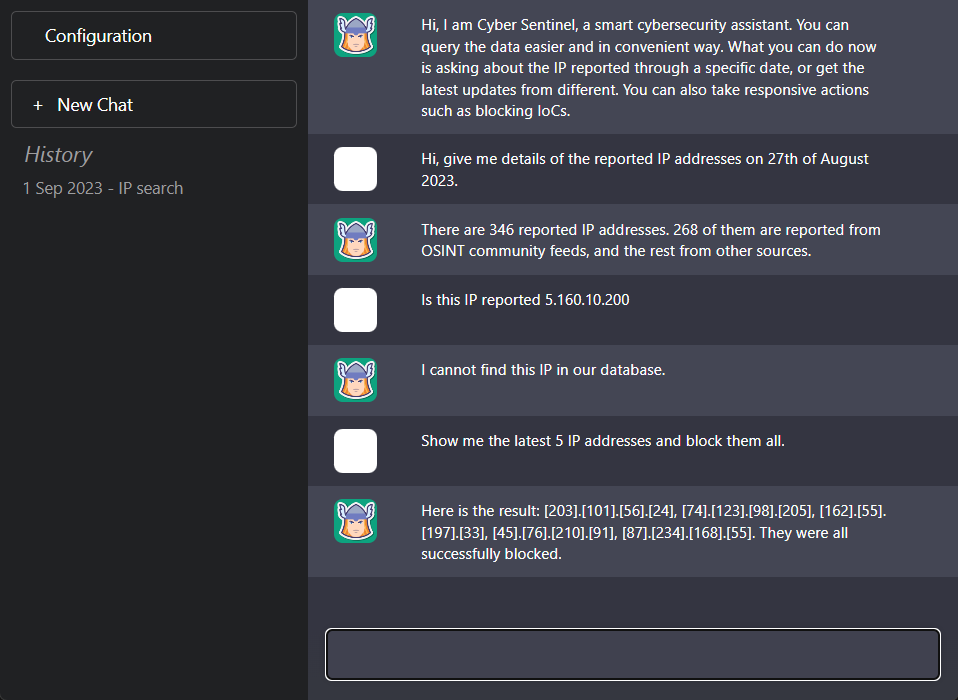}}

\caption{Example screenshot of the user interacting with Cyber Sentinal.}
\label{fig:example_ss}
\end{figure*}

Our experiments show that LLMs can indeed be used for these tasks with minimal effort to some extent, but gauging their precise effectiveness in an operational environment is challenging. In this work, we will implement some basic functionalities regarding each task and discuss the results. Our main emphasis is on showcasing the potential of GenAI in cybersecurity (as shown in Fig. \ref{fig:example_ss}) rather than proposing a novel, end-to-end tool for facilitating cybersecurity pipelines.

The structure of this paper is as follows: Section \ref{sec:relate_work} covers related work and prior research in GenAI and Cybersecurity. Section \ref{sec:methodology} delves into Cyber Sentinel, an approach to leverage conversational agents in cybersecurity vericals. Section \ref{sec:discussion} presents findings from the experiments, their potential implications, and inherent limitations. The paper wraps up in section \ref{sec:conclusion}, summarizing the main points and suggesting directions for future research.

\section{Related Work}
\label{sec:relate_work}
This section will present a literature review, focusing on Generative AI, Network Security, and the intersection of the two.

\subsection{Generative AI}
Generative Artificial Intelligence has emerged as a transformative technology that holds significant promise across various domains, including cybersecurity. Generative AI encompasses a range of techniques that enable machines to create novel content in different formats, including but not limited to sound, text, or images, and is often indistinguishable from human-generated outputs \cite{lamb2021brief}. This technology is underpinned by deep learning architectures, such as \emph{Generative Adversarial Networks (GANs)} \cite{goodfellow2014generative} and \emph{Variational Autoencoders (VAEs)} \cite{kingma2013auto}, which have demonstrated remarkable proficiency in generating images, text, and even complex data sequences.

The field is not solely focused on image generation though, with another class of generative models being transformers \cite{vaswani2017attention } which target text generation. One of the front-runners of these transformer-based generators is a Large Language Model released by OpenAI called \emph{GPT} \cite{radford2018improving}, with the latest version being GPT-4 \cite{OpenAI2023GPT4TR}. GPT-4 can perform a plethora of language-related tasks, such as language translation, sentiment analysis, and question answering, with minimal task-specific fine-tuning \cite{zhang2023complete}. This remarkable versatility has led to widespread exploration of GPT's applications in fields beyond language, including data augmentation \cite{dai2023chataug}, medicine \cite{NEJMsr2214184, wang2023chatcad} and even law \cite{katz2023gpt}.

Although these models are capable of understanding most questions and situations within their context, they sometimes face difficulties in comprehending ambiguous or highly complex queries due to biases in their understanding. Moreover, in specialized expert fields, the abundance of unique abbreviations exacerbates the models' challenges in understanding context, resulting in imprecise and unproductive responses \cite{liu2023summary}.

One of the approaches to overcome the mentioned limitation of LLMs is called Prompt Engineering, which is the process of crafting effective and precise  instructions (i.e., prompts) when working with language models like GPT. It involves formulating input text in a way that guides the model to generate the desired output. Proper prompt engineering is important to obtain accurate and relevant responses from the model, as it helps to influence the way the model interprets and generates text \cite{white2023prompt}. There has been a great effort from the community towards this direction, resulting in prompt engineering almost becoming its own sub-field within language modeling \cite{wei2022chain,yao2023tree,wang2022self}.

Another important aspect is tracking the current user state in a task-oriented dialog context. \emph{Dialog State Tracking} (DST) \cite{williams2016dialog} provides a control over a set of slots which should be given to conversational agents and also facilitate the interaction with other components. A \emph{"slot"} refers to a designated and structured data field within a conversation or input. These slots are utilized to extract specific pieces of information, such as user interests, preferences, or parameters, to facilitate effective communication and task execution. The most important slot in a conversation is \emph{"intent"}, which is the general theme of the conversation that the user's request mainly revolves around. DST should accumulate adequate information during the conversation with AI and is responsible for keeping dialog state and slot values updated at each turn of the conversation.

Among the approaches that address DST, there are some that take slots into account independently. An approach to capture such intricate relationships involves the application of self-attention mechanisms \cite{ye2021slot}. Rather than relying solely on automatic relationship learning, an alternative research avenue leverages the existing knowledge inherent in domain ideas. One compelling strategy involves harnessing the hierarchical structure present in these ontologies, as demonstrated by \cite{li-etal-2021-generation}.

\subsection{Network Security}
Network security stands as a cornerstone of modern cybersecurity strategies, aiming to protect critical data and systems from unauthorized access, attacks, and data breaches. With the evolving threat landscape, traditional security measures have been supplemented by advanced technologies and methodologies. In recent literature, various approaches have been explored to enhance network security. Intrusion Detection Systems (IDS) and Intrusion Prevention Systems (IPS) have been widely studied as means to detect and prevent unauthorized access and malicious activities \cite{khraisat2019survey,azeez2020intrusion}. Machine learning algorithms, such as anomaly detection and behavioral analysis, are also gaining traction in network security due to their ability to identify patterns indicative of suspicious behavior \cite{amrollahi2020enhancing,arevalo2019survey,hou2019use}.

Security Information and Event Management (SIEM)  systems have emerged as essential tools for monitoring and managing security-related events within an organization's IT infrastructure. These systems collect and analyze vast amounts of security data from different sources, including network devices, servers, applications, and user activity logs. SIEMs offer real-time monitoring, threat detection, incident response, and compliance reporting, thereby enabling security teams to identify and respond to potential threats promptly \cite{s21144759}.

Recent research focuses on enhancing SIEM capabilities through integration with machine learning and AI techniques. This fusion enables SIEMs to more accurately identify abnormal behavior and potential security incidents, reducing false positives and enhancing the efficiency of incident response \cite{10.5555/3507788.3507803,doriaidentification,muhammad2023integrated}.

Cyber Threat Intelligence (CTI) plays a crucial role in proactive cybersecurity measures by providing valuable insights into emerging threats, threat actors, and their tactics, techniques, and procedures (TTPs). CTI involves collecting, analyzing, and sharing information about cyber threats to empower organizations with the knowledge needed to anticipate and counteract potential attacks \cite{wagner2019cyber}.

Modern applications of CTI include leveraging big data analytics and AI-driven techniques to process and analyze vast amounts of threat data in real-time \cite{mittal2019cyber,suryotrisongko2022robust}. Automated threat intelligence platforms assist security teams in rapidly identifying and assessing new threats, facilitating quick decision-making and proactive defense measures.

\subsection{LLMs and Cybersecurity}
AI's ability to generalize has effectively replaced traditional rule-based methods with smarter technology \cite{renaud2023}. Nevertheless, the evolving digital environment is not only enhancing technology but also increasing the complexity of cyber threat actors. In the past, cyberspace dealt with relatively basic intrusion attempts, albeit in large numbers; however, the advent of AI-empowered cyberattacks initiated a completely new era, introducing both familiar and unfamiliar changes to cyberattack methods \cite{renaud2023,aryal2021survey}. Thus, one could argue that the advancement of GenAI tools in cybersecurity serves as a double-edged sword, aiding both the defenders and the adversaries \cite{gupta2023chatgpt}.

On the offensive front, LLMs and chatbots (especially ChatGPT) have been used on a range of tasks such as malware development \cite{mckee2022chatbots}, phishing attacks \cite{karanjai2022targeted}, misinformation \cite{withsecure2023} and even false data injection targeting networks \cite{dk_thesis} and industrial systems \cite{Al-Hawawreh2023}. An astute security expert might point out that most of these LLM-based approaches are neither novel nor efficient in achieving their goal. An important point worth mentioning here would be that information regarding cyber offenses involving malicious actions is generally prohibited in many jurisdictions due to legal and ethical considerations, which restricts its accessibility. The availability of large language models like ChatGPT can potentially ease the scarcity of resources for individuals with limited knowledge or skills seeking to bypass ethical constraints when engaging in cyber offenses. Since these LLMs offer a vast amount of information in one location, they can easily provide the comprehensive data necessary to execute various cyber offenses.  

On the other hand, LLMs have been just as effective in helping security experts defend against this new wave of cyber threats. Ransomware mitigation \cite{MCINTOSH2023103424}, synthetic CTI generation \cite{9534192} and intrusion detection \cite{markevych2023review} are all examples of defensive use cases LLMs have had so far. Chatbots are particularly of interest lately, with some works focusing on promoting security awareness \cite{fung2022chatbot} or even finding system vulnerabilities \cite{sanchezintent}. Perhaps the closest use case to our work is \emph{ChatIDS} \cite{juttner2023chatids}, which is a chat-based AI designed to explain IDS alerts to non-experts by using large language models.

\section{Methodology}
\label{sec:methodology}
In this section, we will discuss our proposed novel framework. Subsection \ref{subsec:components} will introduce Cyber Sentinel and give an overview of the other components it interacts with. Subsection \ref{subsec:cyber_sentinel} goes into finer details of the conversational agent itself and discusses how it is prompted by the user and how it communicates with other components. 

\subsection{Components}
\label{subsec:components}
Our proposed framework has four components, all of which are outlined in Fig. \ref{fig:cyber_sentinel_overview} and discussed in detail below:

\begin{figure}[htbp]
\centerline{\includegraphics[width=0.5\textwidth]{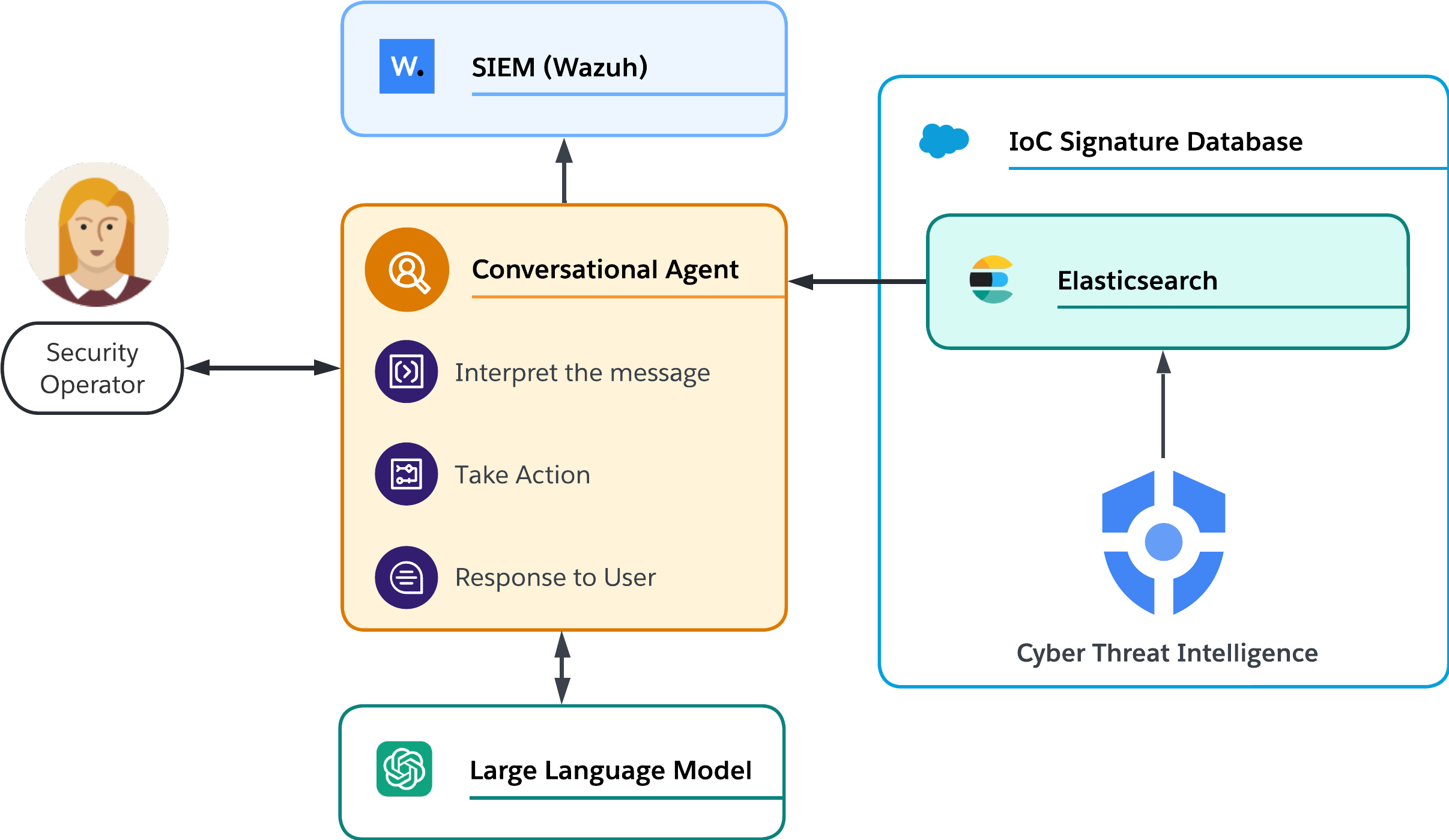}}

\caption{An overview of Cyber Sentinel and components it interacts with.}
\label{fig:cyber_sentinel_overview}
\end{figure}

\subsubsection{IoC Signature Database}
Our CTI module is comprised of an Elasticsearch cluster which is continuously updated by a stream of Indicators of Compromise (IoCs) coming from OSINT data sources. These OSINT feeds are widely used by both academic and industrial cybersecurity communities. Table \ref{tab:osint_sources} includes all of the OSINT sources that were used in our work. 

Elasticsearch offers an extensive API for integration with other components, which is one of the reasons it is commonly used by the cybersecurity community. Whenever a user inquires about any CTI-related resources (e.g. "IoCs reported in the last week"), Cyber Sentinel constructs the appropriate query and then executes that query on the Elasticsearch database through its API.

\begin{table}[]
\centering
\begin{tabular}{|c|c|}
\hline
\textbf{Source Name} & \textbf{Website URL}                                                    \\ \hline
Abuse URL            & https://urlhaus-api.abuse.ch/\\ \hline
Abuese Malware       & https://urlhaus-api.abuse.ch/\\ \hline
Malware Bazaar       &https://bazaar.abuse.ch/\\ \hline
AlienVault           & https://otx.alienvault.com/\\ \hline
Anomali              & https://www.anomali.com/\\ \hline
\end{tabular}
\caption{List of OSINT threat feeds used for Cyber Sentinel in this paper.}
\label{tab:osint_sources}
\end{table}

\subsubsection{Conversational Agent}
The conversational agent is the core part of the framework, integrating every other component together, and also acts as a proxy between system components and the user. It embodies a search engine architecture distinguished by a state-of-the-art, AI-driven interactive response mechanism backed by an LLM. This innovative design empowers users to interact with the software using natural language, formulating their queries in a conversational manner. For instance, a user might inquire, \textit{"Can you display the recent updates from the IP addresses reported on the last day?"}. In reply, the system adeptly compiles and presents relevant data to the user in an intuitive and user-friendly format.

\subsubsection{SIEM}
The SIEM module, empowered by Wazuh, assumes a pivotal role in enhancing the overall security of the system. SIEM delivers a comprehensive array of functionalities, empowering systems to not only detect and monitor potential cybersecurity threats but also to effectively respond to them. Wazuh stands as an open-source security platform renowned for its all-encompassing security monitoring, threat detection, and incident response capabilities. Crafted for profound extensibility and tolerability, Wazuh seamlessly melds a robust SIEM system with an advanced Intrusion Detection System (IDS). The platform centrally accumulates and integrates security-related data originating from various sources including logs, events, and network traffic, thoroughly looking out for potential cybersecurity risks. Wazuh is also capable of fusing real-time threat identification with proactive and reactive response mechanisms, rapidly countering security incidents and eliminating potential threats.

Wazuh's design is based on a server-client architecture, with the main components being Wazuh Manager (server) and Wazuh Agent (client). A typical Wazuh setup includes one (or more in distributed setups) Wazuh manager which collects and analyzes security data from various sources, such as logs, events, and alerts, generated by multiple agents installed on monitored systems. Agents are also capable of taking actions such as updating firewalls on the monitored nodes when instructed by the manager. Optionally, the manager can be complemented with \emph{Elasticsearch} and \emph{Kibana} to index and visualize the collected data.

One of the main use cases of Wazuh is intrusion detection, which is handled through alerts. Alerts are notifications generated by Wazuh when it detects specific security events or suspicious activities on monitored systems. Wazuh uses rules to analyze security-related data, including log files, system events, and configuration changes. These rules define patterns and conditions that, when met, trigger alerts. Wazuh also assigns severity levels to alerts which indicates the potential impact and importance of the detected event, and these could range from \emph{Low} threat levels up to \emph{Critical} levels. In addition to alerting, Wazuh can be configured to trigger response actions when certain alerts occur. These actions can include executing scripts, blocking IP addresses, or other mitigation measures to address security breaches based on threat level.

Wazuh comes with a wide set of predefined rules and responses suitable for most use cases, but it also has the option to define customized rules. These custom rules can be set to check all incoming traffic against a blacklist of IoCs stored in Wazuh's Centralized Database (CDB). As shown in Fig. 1 this capability was used in our work to create custom notifications based on user requests. In other words, security operators can update this CDB through conversation with Cyber Sentinel, and custom rules, then check all traffic in a network against the blacklist and generate alerts if any traffic matches the defined rules. Cyber Sentinel can also be instructed to update Wazuh response actions, for instance, block some of the indicated IPs.

\subsubsection{Large Language Model}
The LLM module serves as the backbone component of the entire system, facilitating the conversational agent in executing its tasks effectively. The LLM that is used in our work is GPT-4 \cite{OpenAI2023GPT4TR}, the latest LLM released by OpenAI. Access to GPT-4 is provided through OpenAI APIs \cite{openai_api}.

\subsection{Cyber Sentinel}
\label{subsec:cyber_sentinel}
This section provides a closer look at Cyber Sentinel, the conversational agent introduced in the previous segment, and its implementation. Example user interactions will also be presented to demonstrate its capabilities.

\begin{figure}[htbp]
\centerline{\includegraphics[width=0.5\textwidth]{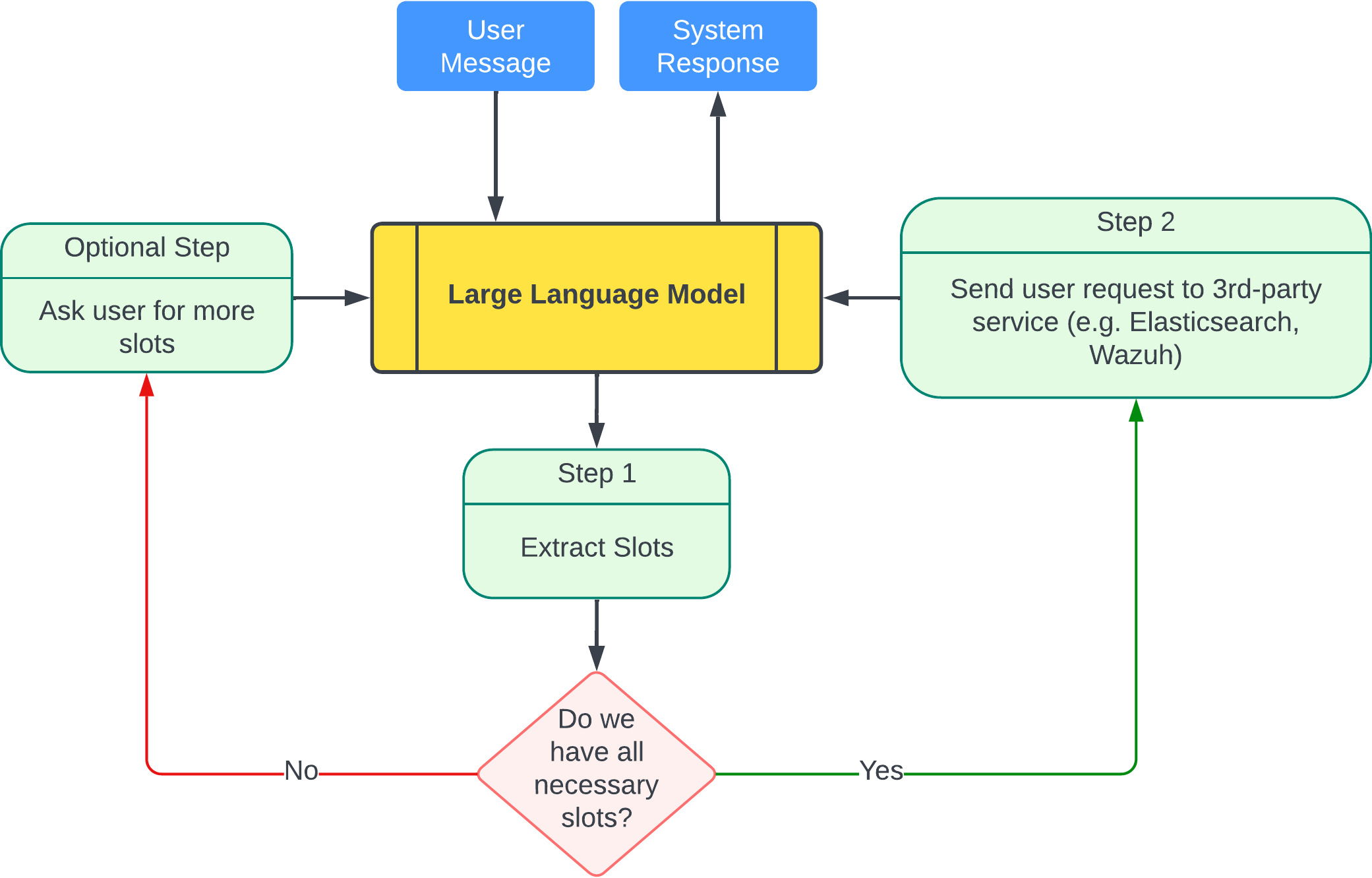}}

\caption{Overview of Cyber Sentinel logic flow. When a user gives the first instruction, all provided slots, including the general intent, are extracted. If 
any of the slots required by the intent have not been provided by the user, Cyber Sentinel prompts the user for the missing slots. Otherwise, Cyber Sentinel proceeds to the second step, which is performing the required action by the user with any of the other components. Lastly, the result of the action is returned to the user as a system message.}
\label{fig:logic_overview}
\end{figure}

\subsubsection{System Design}
The logical implementation includes two segments: The extraction of user intent and slots and the subsequent execution of the requested actions. As defined before, \emph{intents} are the general subject of the user request, and \emph{slots} are supplementary fields for each intent that provide further details about the request. Table \ref{tab:intents_list} contains the list of intents we implemented for Cyber Sentinel to handle with some examples for each intent, and table \ref{tab:intent_recognition} has the list of slots extracted along with intents. These lists are by far non-exhaustive and are only considered as proof-of-concept for our work.

We embraced a sequential approach while utilizing LLMs, chaining GPT-4 models to facilitate user interactions and streamline command processing which is further explained in subsection \ref{subsec:sequential_gpt_models}. GPT-4 models are fed a series of chat messages annotated with distinct dialogue roles user, assistant, and system. Through extensive pre-training and a provided system message, the LLM acquires a profound grasp of its designated role and associated responsibilities, thus enabling it to execute generic question-answering tasks in addition to specific security-related downstream tasks. Fig. \ref{fig:logic_overview} depicts Cyber Sentinel's logic flow diagram, and Fig. \ref{fig:system_template} includes the system template message that is fed to LLM alongside message history upon each interaction.

\begin{table*}[ht]
\centering
\begin{tabular}{|c|c|}
\hline
\textbf{Category} & \textbf{Example} \\
\hline
\multirow{2}{*}{General Question} & Where is the capital of Finland? \\
& How is the weather \\
\hline
\multirow{2}{*}{Cyber security Question} & What is Phishing? \\
& How do banks protect customer data from cyber threats? \\
\hline
\multirow{6}{*}{Query} & Give the latest IP addresses reported in the last 24 hours. \\
& Show the statistics of the latest IoCs in the last 7 days. \\
& Is this email address malicious: \emph{John.Doe@testcompany.com} \\
& Is this URL \emph{John.Doe.com} secure? \\
& Show me all attacks targeting TCP port 9000. \\
& How many attacks reported within the last 24 hours targeting TCP port 23? \\
\hline
\multirow{2}{*}{Action} & Block the IP addresses within subnet \emph{54.12.0.0/16} \\
& Block the hash signature \emph{530ac4...} \\
\hline
\multirow{3}{*}{Hybrid} & Block 130.231.4.98 if it is malicious. \\
& Block all IP addresses reported today. \\
\hline
\end{tabular}
\caption{List of user intentions including user queries and actions to the Wazuh agent}
\label{tab:intents_list}
\end{table*}

\begin{figure}[htbp]
\begin{footnotesize}

You are a cyber security assistant AI called Cyber Sentinel, designed to answer user questions related to cybersecurity and computer security.

In general, questions will be in one of these categories:

1. General questions: The user might ask some irrelevant questions such as "How are you?", or "What is the weather". If the user asks general questions, you should remind them that you are only designed to answer security-related questions. Put \textbf{irrelevant} at the beginning of your response.

2. Cybersecurity and computer security questions: Give users as many details as you can regarding their inquiries. Put \textbf{cybersecurity} at the beginning of your message.

3. Queries: Imagine you have access to a cyber threat intelligence database, in which a number of reported IP addresses, file hashes, and malicious email addresses. Users might want to inquire whether an IP address is reported in the database or not, or have similar queries such as "Show me all reports regarding TCP port 2300", or " Give me the latest updates in the last 24 hours". If you feel that user request is similar to this category, provide \textbf{query} at the beginning of your message.

4. Actions: Lastly, user might ask you to take a security action like blocking an IP address in their system. In this case, they will say something like "Block the ip 127.0.0.1". If you think the user is requesting an action, put \textbf{action} at the beginning of your message.

Keep in mind that user messages might be somewhat different than the examples I provided so be as general as possible. 
\end{footnotesize}

\caption{System instruction prompt in order to instruct GPT-4 to extract intents from user messages.}
\label{fig:system_template}
\end{figure}

\begin{table*}[ht]
    \centering
    \begin{tabular}{|c|c|c|}
        \hline
        \textbf{Slot} & \textbf{Values} & \textbf{Description} \\
        \hline
        Intent & Status, Search, Block, Unblock & Main intention(s) of the request \\
        \hline
        Signature\_Type & IP, Subnet, Email, Hash, URL, Port & Type of requested signature\\
        \hline
        Signature\_Value & String or Number (IPv4,IPv6,email, etc.)  & Signature itself \\
        \hline
        From\_Date & Datetime: 2023/01/01 & Filter search to start from a specific date and time \\
        \hline
        To\_Date & Datetime: 2023/01/02 & Filter search to end on a specific date and time\\
        \hline
        Quantity & Number & Search for the quantity requested \\
        \hline
    \end{tabular}
    \caption{Intent recognition: List of slots extracted from user message}
    \label{tab:intent_recognition}
\end{table*}

\subsubsection{Sequential GPT models}
\label{subsec:sequential_gpt_models}
As discussed previously, we have incorporated a sequential approach in using LLMs in our framework by chaining GPT models to one another. The previous subsection discussed the general architecture and logic of Cyber Sentinel, with Fig. \ref{fig:logic_overview} giving an overview of how it typically interacts with users. In a practical setting, the design exhibits greater complexity than depicted in Fig. \ref{fig:logic_overview}. The procedural steps can be delineated as follows:
\begin{enumerate}
    \item First, we feed user input into GPT-4 alongside the system message template that was shown in Fig. \ref{fig:system_template}. GPT-4 model will label the response with one of the mentioned \emph{general intents} to specify the user's intent. For irrelevant and general questions, we just let GPT provide the answer and pass it back to the user. For the query and action intents, we proceed to step 2.
    \item At this juncture, LLMs are sequentially chained by forwarding the user's message to GPT-4 once more. However, on this occasion, it utilizes a distinct system template message tailored to extract the requisite steps for the desired intent. For this part, we also use a number of prompt engineering techniques such as Chain-of-Thought \cite{wei2022chain} and Self-Consistency \cite{wang2022self} in our system template to further improve GPT-4's reasoning capabilities. The provided template also takes in a number of parameters for dynamically calculating some slots, for instance, when LLM is instructed to create time-based queries, it automatically calculates the time ranges itself. Next, we pass each extracted step to another LLM (GPT-4 again) with a different prompt, this time asking it to extract slots based on the intent. Thanks to the high inference power of GPT-4, we are also able to discern some slots implicitly from user messages. For example, if the user asks for the last 24 hours of activity, GPT-4 is able to infer To\_Date query parameter is the current date and time. Finally, if the result of this step is satisfactory (i.e., all required slots are extracted/inferred and the action can be performed), we proceed to the next step. Otherwise, we prompt the user to provide further details regarding the action they would like to take.
    \item Finally, when both intent and related slots are extracted, we perform the requested action on either the Elasticsearch database or the Wazuh cluster through their API.
\end{enumerate}

\begin{figure}[htbp]
\centerline{\includegraphics[width=0.4\textwidth]{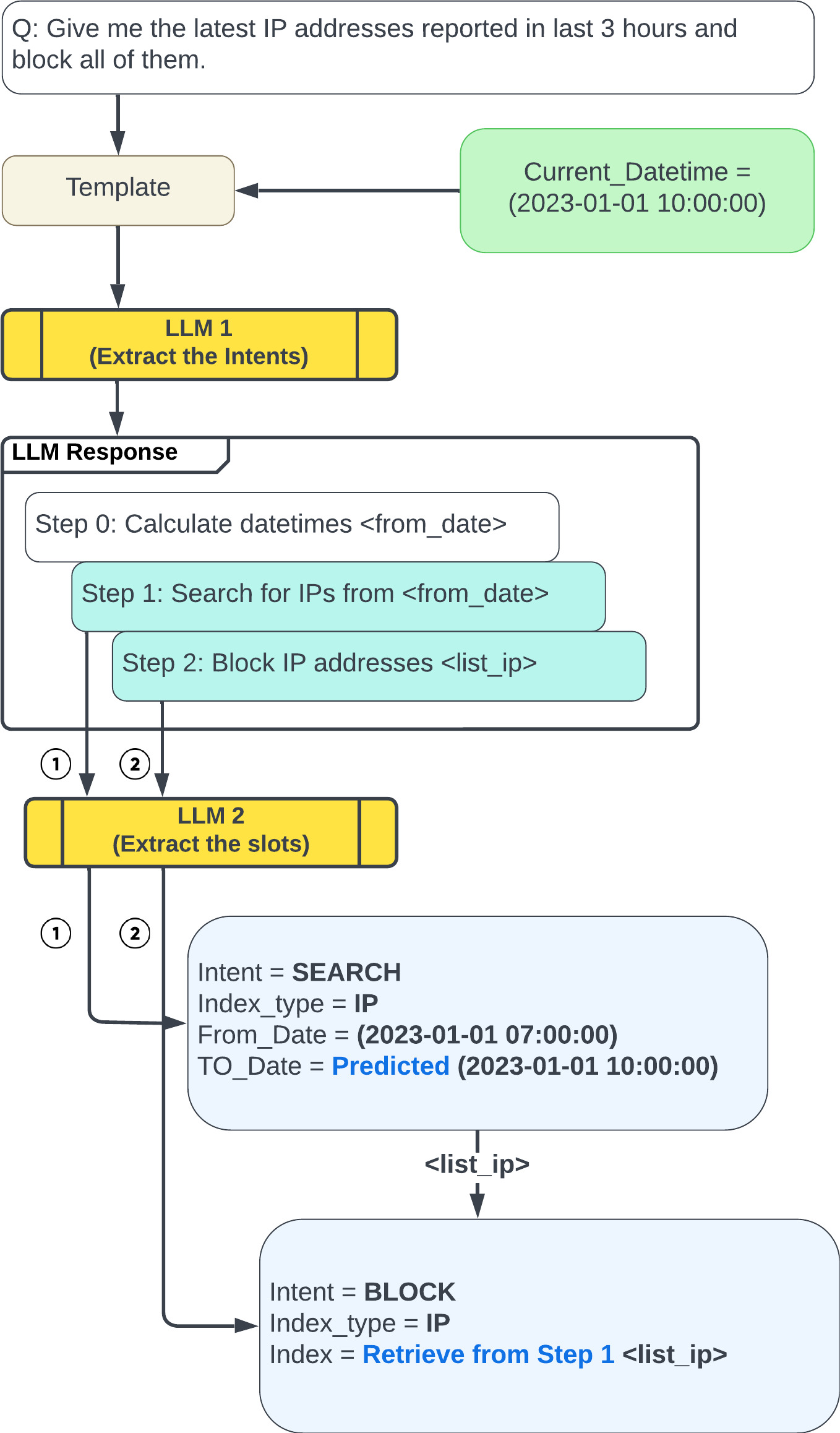}}

\caption{A sample of Cyber Sentinel's chaining process. First, LLM is prompted with a user message and a template dynamically created with the current date and time. First LLM constructs the necessary steps needed to perform the requested action through Chain-of-Thought, and extracts/formulates related \emph{parameters} (e.g. from\_date, list\_ip). Afterward, each of these generated steps is passed to another LLM which extracts given slots (like intent, index\_type, from\_date, etc.) and fills in other implicit slots (such as predicted to\_date and retrieved list\_ip).}
\label{fig:example_LLM_chaining}
\end{figure}

An example of this process is depicted in Fig. \ref{fig:example_LLM_chaining}.

\section{Discussion}
\label{sec:discussion}
The preceding section presented the framework and elucidated the functionality and integration of its various components. In this segment, the achieved results will be discussed, alongside their implications for the cybersecurity community. Potential applications, inclusive of misuse scenarios for the tool, will also be explored. The discussion will conclude by highlighting some inherent limitations of this research

\subsection{Impacts and use cases}
The integration of conversational agents like Cyber Sentinel into cybersecurity operations has the potential to yield a range of significant impacts, ultimately enhancing the effectiveness and efficiency of security measures. Below, the advantages are explored in detail, highlighting how these agents can streamline operational processes, facilitate rapid detection and response to security events, and contribute to a more robust and adaptive security posture.

\begin{itemize}
    \item \textbf{Improved Threat Detection and Response:} One of the primary impacts of Cyber Sentinel is its capacity to significantly improve threat detection and response times. By leveraging its cyber threat intelligence module, the system can continuously monitor and analyze vast datasets of security events and alerts from an SIEM. Through natural language interactions with security analysts, it can swiftly pinpoint anomalies, recognize patterns, and identify potential security breaches. Additionally, it can take actions based on this intelligence with minimal instructions provided by security analysts, thus reducing the window of vulnerability and minimizing potential damages to systems.

\item \textbf{Enhanced Operational Efficiency:} The introduction of a conversational agent like Cyber Sentinel can streamline security operations. Security analysts often face a deluge of alerts, leading to alert fatigue and potentially missing critical threats. Cyber Sentinel's ability to prioritize and present relevant information to analysts in a conversational format alleviates this challenge. It can automate routine tasks such as querying SIEM data or updating firewall rules based on predefined policies. This automation not only reduces manual workload but also ensures consistency and accuracy in security responses.

\item \textbf{Real-time Collaboration and Knowledge Sharing:} Cyber Sentinel's conversational capabilities facilitate real-time collaboration among security team members. Analysts can engage with the agent to discuss ongoing incidents, share insights, and collectively make decisions. The agent serves as a central hub for disseminating threat intelligence and best practices, promoting knowledge sharing among team members. This fosters a more collaborative and responsive security environment.
\end{itemize}

The deployment of Cyber Sentinel in cybersecurity operations opens up a spectrum of practical use cases, each showcasing its adaptability and value across diverse scenarios:
\begin{itemize}
    \item \textbf{Incident Triage and Investigation:} In the context of incident response, Cyber Sentinel can play a crucial role in triaging and investigating security incidents. It can gather initial information from the SIEM, classify the incident severity, and provide analysts with a starting point for their investigations. This allows security teams to focus their efforts on high-priority incidents while the agent handles routine tasks.

    \item \textbf{Threat Intelligence Analysis:} The agent's cyber threat intelligence module empowers organizations to stay proactive in the face of evolving threats. It continuously monitors external threat feeds, aggregating and analyzing data from various sources. It can then provide timely intelligence updates to security analysts, allowing them to adjust their defenses and strategies accordingly.

    \item \textbf{Security Policy Management:} Cyber Sentinel can also streamline security policy management. Security policies are often complex, and keeping firewall rules up-to-date can be challenging. The agent can assist in the process by reviewing policy changes, suggesting updates based on threat intelligence, and implementing approved changes. This ensures that security policies remain agile and aligned with emerging threats.

    \item \textbf{Security Awareness and Training:} Beyond its operational roles, Cyber Sentinel can contribute to improving the overall security posture of an organization. It can serve as an educational tool by providing security awareness training to employees. Through interactive conversations, it can impart knowledge about common threats, safe online practices, and company-specific security policies.

\end{itemize}

While the potential benefits of Cyber Sentinel in cybersecurity are substantial, it is essential to acknowledge the potential for misuse or malicious purposes. Conversational agents like Cyber Sentinel, if compromised or manipulated, could be used to disseminate false information, disrupt in-place security measures, or even automate cyberattacks. This is particularly of importance as methods to \emph{jailbreak} LLMs (i.e., exploiting these models to enable them to perform tasks or generate content that goes beyond their intended use cases) keep getting more sophisticated each day \cite{gupta2023chatgpt}. Therefore, strict access controls, authentication mechanisms, and continuous monitoring are necessary to prevent unauthorized access or manipulation of the agent.

Additionally, ethical considerations come into play when designing and deploying such agents. Ensuring that the agent respects privacy and complies with relevant regulations is paramount. Furthermore, measures should be in place to prevent the unintended disclosure of sensitive information during interactions with the conversational agent.

In conclusion, the integration of conversational agents like Cyber Sentinel in cybersecurity operations brings forth significant impacts and a multitude of practical use cases. These agents, by harnessing the capabilities of artificial intelligence and natural language processing, have the potential to revolutionize how organizations defend against cyber threats, offering a promising path towards more robust and efficient cybersecurity strategies.

\subsection{Limitations}
Despite the great impacts and numerous use cases that were mentioned in the previous subsection, Cyber Sentinel also has a number of limitations that need to be considered for further work:
\begin{itemize}
    \item \textbf{Human Oversight and Decision-Making:} Cybersecurity tasks often require human expertise and judgment, particularly in situations involving complex, nuanced, or novel threats. Relying entirely on the agent without human oversight can be risky, as it may not always make the most appropriate decisions. This is especially true in cases where existing intelligence is not of good enough quality for the agent to be taken advantage of, or when faced with sophisticated threats and zero-day attacks.
    \item \textbf{Privacy Concerns and Ethical Issues:} The deployment of a conversational agent within a cybersecurity ecosystem raises security and privacy concerns. If not adequately secured, the agent itself could become a target for attackers seeking to manipulate its responses or gather sensitive information. Furthermore, conversational agents are susceptible to biases present in their training data just like other AI systems. This could lead to biased recommendations or responses, potentially causing ethical dilemmas or reinforcing existing biases within security practices.
    \item \textbf{Regulatory Compliance and Resource Intensity:} Compliance with cybersecurity regulations and standards is crucial for organizations. Implementing conversational agents must align with regulatory requirements, which can add complexity to the deployment process. Additionally, the deployment and maintenance of a conversational agent like Cyber Sentinel can be resource-intensive as LLMs are notoriously known to require a significant amount of processing power. Organizations must allocate sufficient resources for initial setup, ongoing training, and monitoring to ensure the agent operates effectively.
\end{itemize}
\section{Conclusions}
In this paper, we introduced a conversational agent called \emph{Cyber Sentinel}, which can be used for streamlining cyber security. Cyber Sentinel is capable of helping security analysts perform a range of security-related tasks, from querying a cyber threat intelligence feed to managing an SIEM's configuration. We went over how Cyber Sentinel works and how it interacts with other components in a typical cybersecurity organization. The potential impacts and limitations of the tool were also briefly considered. 

It should be mentioned once again that this is a work in progress and only presented now as a proof of concept that conversational agents and LLMs can be immensely powerful in certain cybersecurity tasks. Nevertheless, many steps remain to be explored before AI is ready to step into the role of security analyst itself. Some possible directions for future work may be:
\begin{itemize}
    \item \textbf{Adaptive Threat Detection:} Investigate the development of conversational agents with self-learning capabilities to adapt to evolving threats. This research could explore techniques such as reinforcement learning to enable agents to continually improve their threat detection accuracy.
    \item \textbf{Explainability and Actionability:} Explore methods for making conversational agents more transparent and interpretable. Further research into explainable AI \cite{10.1007/978-3-030-32236-6_51} and actionable AI \cite{10.1145/3424690} techniques could help build trust and confidence in AI-driven security decision-making.
    \item \textbf{Multi-Agent Systems:} Study the potential of multi-agent systems where conversational agents work collaboratively to enhance security operations. Research could explore how these agents can distribute tasks, share insights, and collectively respond to complex cyber threats.
    \item \textbf{Ethical Hacking and Vulnerability Assessment:} Research the application of conversational agents in ethical hacking and vulnerability assessment, including automated penetration testing and vulnerability scanning. Some work \cite{happe2023getting,deng2023pentestgpt} have already started down this path but further investigation is still required.
\end{itemize}
\label{sec:conclusion}

\bibliographystyle{ieeetr}
\bibliography{citations}

\begin{thebibliography}{10}

\bibitem{10.1145/3382025.3414952}
A.~J. Varela-Vaca, R.~M. Gasca, J.~A. Carmona-Fombella, and M.~T. G\'{o}mez-L\'{o}pez, ``Amadeus: Towards the automated security testing,'' in {\em Proceedings of the 24th ACM Conference on Systems and Software Product Line: Volume A - Volume A}, SPLC '20, (New York, NY, USA), Association for Computing Machinery, 2020.

\bibitem{8908835}
S.~Liu, Z.~Hou, Y.~Guo, and L.~Guo, ``A novel modified robust model-free adaptive control method for a class of nonlinear systems with time delay,'' in {\em 2019 IEEE 8th Data Driven Control and Learning Systems Conference (DDCLS)}, pp.~1329--1334, 2019.

\bibitem{9458828}
P.~Gao, F.~Shao, X.~Liu, X.~Xiao, Z.~Qin, F.~Xu, P.~Mittal, S.~R. Kulkarni, and D.~Song, ``Enabling efficient cyber threat hunting with cyber threat intelligence,'' in {\em 2021 IEEE 37th International Conference on Data Engineering (ICDE)}, pp.~193--204, 2021.

\bibitem{10224990b}
S.~Sheikhi and P.~Kostakos, ``Cyber threat hunting using unsupervised federated learning and adversary emulation,'' in {\em 2023 IEEE International Conference on Cyber Security and Resilience (CSR)}, pp.~315--320, 2023.

\bibitem{Ansari2022}
M.~A. Ali~Ansari, M.~Shahid, A.~J. Khan, M.~S.~e. Ali, and K.~A. Mehmood, ``The effects of project assessment, safety management training, and risk assessment on migrant labor,'' {\em Journal of Social Sciences Review}, vol.~2, p.~35–44, Dec. 2022.

\bibitem{8993037}
L.~Wang and R.~Jones, ``Big data analytics in cybersecurity: Network data and intrusion prediction,'' in {\em 2019 IEEE 10th Annual Ubiquitous Computing, Electronics \& Mobile Communication Conference (UEMCON)}, pp.~0105--0111, 2019.

\bibitem{sheikhi2022novel}
S.~Sheikhi and P.~Kostakos, ``A novel anomaly-based intrusion detection model using psogwo-optimized bp neural network and ga-based feature selection,'' {\em Sensors}, vol.~22, no.~23, p.~9318, 2022.

\bibitem{9527966}
O.~Ajayi and A.~Gangopadhyay, ``Dahid: Domain adaptive host-based intrusion detection,'' in {\em 2021 IEEE International Conference on Cyber Security and Resilience (CSR)}, pp.~467--472, 2021.

\bibitem{CHORAS2021705}
M.~Choraś and M.~Pawlicki, ``Intrusion detection approach based on optimised artificial neural network,'' {\em Neurocomputing}, vol.~452, pp.~705--715, 2021.

\bibitem{s19051114}
L.~Fernández~Maimó, A.~Huertas~Celdrán, A.~L. Perales~Gómez, F.~J. García~Clemente, J.~Weimer, and I.~Lee, ``Intelligent and dynamic ransomware spread detection and mitigation in integrated clinical environments,'' {\em Sensors}, vol.~19, no.~5, 2019.

\bibitem{9406576}
P.~Nespoli, F.~G. Mármol, and J.~M. Vidal, ``A bio-inspired reaction against cyberattacks: Ais-powered optimal countermeasures selection,'' {\em IEEE Access}, vol.~9, pp.~60971--60996, 2021.

\bibitem{cao2023comprehensive}
Y.~Cao, S.~Li, Y.~Liu, Z.~Yan, Y.~Dai, P.~S. Yu, and L.~Sun, ``A comprehensive survey of ai-generated content (aigc): A history of generative ai from gan to chatgpt,'' {\em arXiv preprint arXiv:2303.04226}, 2023.

\bibitem{devlin2018bert}
J.~Devlin, M.-W. Chang, K.~Lee, and K.~Toutanova, ``Bert: Pre-training of deep bidirectional transformers for language understanding,'' {\em arXiv preprint arXiv:1810.04805}, 2018.

\bibitem{setianto2021gpt}
F.~Setianto, E.~Tsani, F.~Sadiq, G.~Domalis, D.~Tsakalidis, and P.~Kostakos, ``Gpt-2c: A parser for honeypot logs using large pre-trained language models,'' in {\em Proceedings of the 2021 IEEE/ACM International Conference on Advances in Social Networks Analysis and Mining}, pp.~649--653, 2021.

\bibitem{radford2018improving}
A.~Radford, K.~Narasimhan, T.~Salimans, I.~Sutskever, {\em et~al.}, ``Improving language understanding by generative pre-training,'' {\em OpenAI}, 2018.

\bibitem{10.1007/978-3-030-32236-6_51}
F.~Xu, H.~Uszkoreit, Y.~Du, W.~Fan, D.~Zhao, and J.~Zhu, ``Explainable ai: A brief survey on history, research areas, approaches and challenges,'' in {\em Natural Language Processing and Chinese Computing} (J.~Tang, M.-Y. Kan, D.~Zhao, S.~Li, and H.~Zan, eds.), (Cham), pp.~563--574, Springer International Publishing, 2019.

\bibitem{10.1145/3424690}
D.~Wilkins, S.~Lindley, M.~Meijer, R.~Banks, and B.~Burlin, ``Designing ai systems that make organizational knowledge actionable,'' {\em Interactions}, vol.~27, p.~72–75, nov 2020.

\bibitem{linkov2020cybertrust}
I.~Linkov, S.~Galaitsi, B.~D. Trump, J.~M. Keisler, and A.~Kott, ``Cybertrust: From explainable to actionable and interpretable artificial intelligence,'' {\em Computer}, vol.~53, no.~9, pp.~91--96, 2020.

\bibitem{lamb2021brief}
A.~Lamb, ``A brief introduction to generative models,'' {\em arXiv preprint arXiv:2103.00265}, 2021.

\bibitem{goodfellow2014generative}
I.~Goodfellow, J.~Pouget-Abadie, M.~Mirza, B.~Xu, D.~Warde-Farley, S.~Ozair, A.~Courville, and Y.~Bengio, ``Generative adversarial nets,'' {\em Advances in neural information processing systems}, vol.~27, 2014.

\bibitem{kingma2013auto}
D.~P. Kingma and M.~Welling, ``Auto-encoding variational bayes,'' {\em arXiv preprint arXiv:1312.6114}, 2013.

\bibitem{vaswani2017attention}
A.~Vaswani, N.~Shazeer, N.~Parmar, J.~Uszkoreit, L.~Jones, A.~N. Gomez, {\L}.~Kaiser, and I.~Polosukhin, ``Attention is all you need,'' {\em Advances in neural information processing systems}, vol.~30, 2017.

\bibitem{OpenAI2023GPT4TR}
OpenAI, ``Gpt-4 technical report,'' {\em ArXiv}, vol.~abs/2303.08774, 2023.

\bibitem{zhang2023complete}
C.~Zhang, C.~Zhang, S.~Zheng, Y.~Qiao, C.~Li, M.~Zhang, S.~K. Dam, C.~M. Thwal, Y.~L. Tun, L.~L. Huy, {\em et~al.}, ``A complete survey on generative ai (aigc): Is chatgpt from gpt-4 to gpt-5 all you need?,'' {\em arXiv preprint arXiv:2303.11717}, 2023.

\bibitem{dai2023chataug}
H.~Dai, Z.~Liu, W.~Liao, X.~Huang, Z.~Wu, L.~Zhao, W.~Liu, N.~Liu, S.~Li, D.~Zhu, {\em et~al.}, ``Chataug: Leveraging chatgpt for text data augmentation,'' {\em arXiv preprint arXiv:2302.13007}, 2023.

\bibitem{NEJMsr2214184}
P.~Lee, S.~Bubeck, and J.~Petro, ``Benefits, limits, and risks of gpt-4 as an ai chatbot for medicine,'' {\em New England Journal of Medicine}, vol.~388, no.~13, pp.~1233--1239, 2023.
\newblock PMID: 36988602.

\bibitem{wang2023chatcad}
S.~Wang, Z.~Zhao, X.~Ouyang, Q.~Wang, and D.~Shen, ``Chatcad: Interactive computer-aided diagnosis on medical image using large language models,'' {\em arXiv preprint arXiv:2302.07257}, 2023.

\bibitem{katz2023gpt}
D.~M. Katz, M.~J. Bommarito, S.~Gao, and P.~Arredondo, ``Gpt-4 passes the bar exam,'' {\em Available at SSRN 4389233}, 2023.

\bibitem{liu2023summary}
Y.~Liu, T.~Han, S.~Ma, J.~Zhang, Y.~Yang, J.~Tian, H.~He, A.~Li, M.~He, Z.~Liu, {\em et~al.}, ``Summary of chatgpt/gpt-4 research and perspective towards the future of large language models,'' {\em arXiv preprint arXiv:2304.01852}, 2023.

\bibitem{white2023prompt}
J.~White, Q.~Fu, S.~Hays, M.~Sandborn, C.~Olea, H.~Gilbert, A.~Elnashar, J.~Spencer-Smith, and D.~C. Schmidt, ``A prompt pattern catalog to enhance prompt engineering with chatgpt,'' {\em arXiv preprint arXiv:2302.11382}, 2023.

\bibitem{wei2022chain}
J.~Wei, X.~Wang, D.~Schuurmans, M.~Bosma, F.~Xia, E.~Chi, Q.~V. Le, D.~Zhou, {\em et~al.}, ``Chain-of-thought prompting elicits reasoning in large language models,'' {\em Advances in Neural Information Processing Systems}, vol.~35, pp.~24824--24837, 2022.

\bibitem{yao2023tree}
S.~Yao, D.~Yu, J.~Zhao, I.~Shafran, T.~L. Griffiths, Y.~Cao, and K.~Narasimhan, ``Tree of thoughts: Deliberate problem solving with large language models,'' {\em arXiv preprint arXiv:2305.10601}, 2023.

\bibitem{wang2022self}
X.~Wang, J.~Wei, D.~Schuurmans, Q.~Le, E.~Chi, S.~Narang, A.~Chowdhery, and D.~Zhou, ``Self-consistency improves chain of thought reasoning in language models,'' {\em arXiv preprint arXiv:2203.11171}, 2022.

\bibitem{williams2016dialog}
J.~D. Williams, A.~Raux, and M.~Henderson, ``The dialog state tracking challenge series: A review,'' {\em Dialogue \& Discourse}, vol.~7, no.~3, pp.~4--33, 2016.

\bibitem{ye2021slot}
F.~Ye, J.~Manotumruksa, Q.~Zhang, S.~Li, and E.~Yilmaz, ``Slot self-attentive dialogue state tracking,'' in {\em Proceedings of the Web Conference 2021}, pp.~1598--1608, 2021.

\bibitem{li-etal-2021-generation}
X.~Li, Q.~Li, W.~Wu, and Q.~Yin, ``Generation and extraction combined dialogue state tracking with hierarchical ontology integration,'' in {\em Proceedings of the 2021 Conference on Empirical Methods in Natural Language Processing}, (Online and Punta Cana, Dominican Republic), pp.~2241--2249, Association for Computational Linguistics, Nov. 2021.

\bibitem{khraisat2019survey}
A.~Khraisat, I.~Gondal, P.~Vamplew, and J.~Kamruzzaman, ``Survey of intrusion detection systems: techniques, datasets and challenges,'' {\em Cybersecurity}, vol.~2, no.~1, pp.~1--22, 2019.

\bibitem{azeez2020intrusion}
N.~A. Azeez, T.~M. Bada, S.~Misra, A.~Adewumi, C.~Van~der Vyver, and R.~Ahuja, ``Intrusion detection and prevention systems: an updated review,'' {\em Data Management, Analytics and Innovation: Proceedings of ICDMAI 2019, Volume 1}, pp.~685--696, 2020.

\bibitem{amrollahi2020enhancing}
M.~Amrollahi, S.~Hadayeghparast, H.~Karimipour, F.~Derakhshan, and G.~Srivastava, ``Enhancing network security via machine learning: opportunities and challenges,'' {\em Handbook of big data privacy}, pp.~165--189, 2020.

\bibitem{arevalo2019survey}
J.~Arevalo~Herrera and J.~E. Camargo, ``A survey on machine learning applications for software defined network security,'' in {\em Applied Cryptography and Network Security Workshops: ACNS 2019 Satellite Workshops, SiMLA, Cloud S\&P, AIBlock, and AIoTS, Bogota, Colombia, June 5--7, 2019, Proceedings 17}, pp.~70--93, Springer, 2019.

\bibitem{hou2019use}
S.~Hou and X.~Huang, ``Use of machine learning in detecting network security of edge computing system,'' in {\em 2019 IEEE 4th International Conference on Big Data Analytics (ICBDA)}, pp.~252--256, IEEE, 2019.

\bibitem{s21144759}
G.~González-Granadillo, S.~González-Zarzosa, and R.~Diaz, ``Security information and event management (siem): Analysis, trends, and usage in critical infrastructures,'' {\em Sensors}, vol.~21, no.~14, 2021.

\bibitem{10.5555/3507788.3507803}
S.~M.~M. Hossain, R.~Couturier, J.~Rusk, and K.~B. Kent, ``Automatic event categorizer for siem,'' in {\em Proceedings of the 31st Annual International Conference on Computer Science and Software Engineering}, CASCON '21, (USA), p.~104–112, IBM Corp., 2021.

\bibitem{doriaidentification}
A.~DORIA, {\em Identification of cyber attacks using Next Generation protection tools: NGFW, NG-SIEM, AI and Machine Learning}.
\newblock PhD thesis, University of Padova, 2022.

\bibitem{muhammad2023integrated}
A.~R. Muhammad, P.~Sukarno, and A.~A. Wardana, ``Integrated security information and event management (siem) with intrusion detection system (ids) for live analysis based on machine learning,'' {\em Procedia Computer Science}, vol.~217, pp.~1406--1415, 2023.

\bibitem{wagner2019cyber}
T.~D. Wagner, K.~Mahbub, E.~Palomar, and A.~E. Abdallah, ``Cyber threat intelligence sharing: Survey and research directions,'' {\em Computers \& Security}, vol.~87, p.~101589, 2019.

\bibitem{mittal2019cyber}
S.~Mittal, A.~Joshi, and T.~Finin, ``Cyber-all-intel: An ai for security related threat intelligence,'' {\em arXiv preprint arXiv:1905.02895}, 2019.

\bibitem{suryotrisongko2022robust}
H.~Suryotrisongko, Y.~Musashi, A.~Tsuneda, and K.~Sugitani, ``Robust botnet dga detection: Blending xai and osint for cyber threat intelligence sharing,'' {\em IEEE Access}, vol.~10, pp.~34613--34624, 2022.

\bibitem{renaud2023}
K.~Renaud, M.~Warkentin, and G.~Westerman, {\em From ChatGPT to HackGPT: Meeting the Cybersecurity Threat of Generative AI}.
\newblock California: O’Reilly, 2023.

\bibitem{aryal2021survey}
K.~Aryal, M.~Gupta, and M.~Abdelsalam, ``A survey on adversarial attacks for malware analysis,'' {\em arXiv preprint arXiv:2111.08223}, 2021.

\bibitem{gupta2023chatgpt}
M.~Gupta, C.~Akiri, K.~Aryal, E.~Parker, and L.~Praharaj, ``From chatgpt to threatgpt: Impact of generative ai in cybersecurity and privacy,'' {\em IEEE Access}, 2023.

\bibitem{mckee2022chatbots}
F.~McKee and D.~Noever, ``Chatbots in a botnet world,'' {\em arXiv preprint arXiv:2212.11126}, 2022.

\bibitem{karanjai2022targeted}
R.~Karanjai, ``Targeted phishing campaigns using large scale language models,'' {\em arXiv preprint arXiv:2301.00665}, 2022.

\bibitem{withsecure2023}
A.~Patel and J.~Sattler, ``Creatively malicious prompt engineering.'' \url{https://labs.withsecure.com/publications/creatively-malicious-prompt-engineering}.
\newblock Accessed: 2023-09-01.

\bibitem{dk_thesis}
D.~K. Kholgh, {\em Synthetic network traffic generation using generative modeling}.
\newblock Msc thesis, University of Oulu, Oulu, Finland, 2023.

\bibitem{Al-Hawawreh2023}
M.~Al-Hawawreh, A.~Aljuhani, and Y.~Jararweh, ``Chatgpt for cybersecurity: practical applications, challenges, and future directions,'' {\em Cluster Computing}, Aug 2023.

\bibitem{MCINTOSH2023103424}
T.~McIntosh, T.~Liu, T.~Susnjak, H.~Alavizadeh, A.~Ng, R.~Nowrozy, and P.~Watters, ``Harnessing gpt-4 for generation of cybersecurity grc policies: A focus on ransomware attack mitigation,'' {\em Computers \& Security}, vol.~134, p.~103424, 2023.

\bibitem{9534192}
P.~Ranade, A.~Piplai, S.~Mittal, A.~Joshi, and T.~Finin, ``Generating fake cyber threat intelligence using transformer-based models,'' in {\em 2021 International Joint Conference on Neural Networks (IJCNN)}, pp.~1--9, 2021.

\bibitem{markevych2023review}
M.~Markevych and M.~Dawson, ``A review of enhancing intrusion detection systems for cybersecurity using artificial intelligence (ai),'' in {\em International conference KNOWLEDGE-BASED ORGANIZATION}, vol.~29,3, pp.~30--37, 2023.

\bibitem{fung2022chatbot}
Y.-C. Fung and L.-K. Lee, ``A chatbot for promoting cybersecurity awareness,'' in {\em Cyber Security, Privacy and Networking: Proceedings of ICSPN 2021}, pp.~379--387, Springer, 2022.

\bibitem{sanchezintent}
S.~Sanchez and M.~Franco, ``An intent-aware chatbot for cybersecurity recommendation.''

\bibitem{juttner2023chatids}
V.~J{\"u}ttner, M.~Grimmer, and E.~Buchmann, ``Chatids: Explainable cybersecurity using generative ai,'' {\em arXiv preprint arXiv:2306.14504}, 2023.

\bibitem{openai_api}
OpenAI, ``Openai api reference.'' \url{https://platform.openai.com/docs/api-reference}.
\newblock Accessed: 2023-09-01.

\bibitem{happe2023getting}
A.~Happe and J.~Cito, ``Getting pwn'd by ai: Penetration testing with large language models,'' {\em arXiv preprint arXiv:2308.00121}, 2023.

\bibitem{deng2023pentestgpt}
G.~Deng, Y.~Liu, V.~Mayoral-Vilches, P.~Liu, Y.~Li, Y.~Xu, T.~Zhang, Y.~Liu, M.~Pinzger, and S.~Rass, ``Pentestgpt: An llm-empowered automatic penetration testing tool,'' {\em arXiv preprint arXiv:2308.06782}, 2023.

\end{thebibliography}

\end{document}